\documentclass[preprint,aps,showpacs,nofootinbib,preprintnumbers,amsmath,amssymb]{revtex4-1}
\usepackage{amssymb}
\usepackage{epsfig}
\usepackage{graphicx}% Include figure files
\usepackage{subfigure}
\usepackage{dcolumn}% Align table columns on decimal point
\usepackage{bm}% bold math
\usepackage[usenames ,dvipsnames]{xcolor}
\usepackage{slashed}

\begin{document}

\title{Exothermic Dark Matter with Light Mediator after LUX and PandaX-II in 2016}

\author{Chao-Qiang~Geng$^{1,2,3}$\footnote{geng@phys.nthu.edu.tw}, 
Da~Huang$^{4}$\footnote{dahuang@fuw.edu.pl} and Chun-Hao~Lee$^{2}$\footnote{lee.chunhao9112@gmail.com}
}
 \affiliation{$^{1}$School of Physics and Information Engineering, Shanxi Normal University, Linfen, 041004, China\\
  $^{2}$Department of Physics, National Tsing Hua University, Hsinchu, Taiwan\\
  $^{3}$Physics Division, National Center for Theoretical Sciences, Hsinchu, Taiwan \\
  $^{4}$Faculty of Physics, University of Warsaw, Pasteura 5, 02-093 Warsaw, Poland}

\date{\today}

\begin{abstract}
Dark matter (DM) direct detections are investigated for models with the following properties: isospin-violating couplings, exothermic scatterings, and/or a light mediator, with the aim to reduce the tension between the CDMS-Si positive signals and other negative searches. In particular, we focus on the non-standard effective operators which could lead to the spin-independent DM-nucleus scatterings with non-trivial dependences on the transfer momentum or DM velocity. As a result, such effective operator choices have the very mild effects on the final fittings. Furthermore, by including the latest constraints from LUX, PandaX-II,  XENON1T and PICO-60, we find that, for almost all the considered models, the predicted CDMS-Si signal regions are either severely constrained or completely excluded by the LUX, PandaX-II,  XENON1T and PICO-60 data, including the most promising Xe-phobic exothermic DM models with/without a light mediator. Therefore, we conclude that it is very difficult for the present DM framework to explain the CDMS-Si excess.
\end{abstract}

%%\pacs{95.35.+d, 13.85.Tp, 14.80.-j, 98.70.Sa,}
%\keywords{Dark Matter}
\maketitle

\section{Introduction}
\label{s1}
Dark matter (DM), a term coined by Fritz Zwicky in 1933~\cite{Zwicky:1937zza}, has its existence been proven by many persuading astrophysical and cosmological evidences~\cite{Rubin:1970zza, Jee:2007nx,PDG,Planck}. However, its particle nature is still obscure. Based on many theoretical motivations, the weakly interacting massive particles (WIMPs)~\cite{Steigman:1984ac,Jungman:1995df,Feng:2010gw} are among the leading DM candidates with the mass scale from GeV to TeV. One promising way to search for WIMPs is direct DM detections~\cite{Goodman:1984dc,FormFactor}, which try to measure the nuclear recoil energies in detectors deposited by WIMP collisions. 

Currently, several collaborations, such as DAMA~\cite{Bernabei:2008yi,Bernabei:2010mq}, GoGeNT~\cite{Aalseth:2010vx,Aalseth:2011wp,Aalseth:2012if,Aalseth:2014eft,Aalseth:2014jpa}, CRESST-II~\cite{CRESST-II-S} and CDMS-Si~\cite{CDMS-Si}, have reported potential signals for light WIMPs with mass of $1 \sim 10$~GeV, which are in conflict with the negative results from other experiments, including SuperCDMS~\cite{SuperCDMS_Ge}, CDMSlite~\cite{Agnese:2013jaa, Agnese:2015nto}, XENON10~\cite{XENON10}, XENON100~\cite{XENON100, XENON100a,XENON100b}, LUX~\cite{LUX2013, LUX2015,LUX2016}, CDEX~\cite{CDEX14, CDEX16}, PandaX~\cite{PandaX14, PandaX16a, PandaX16b}, PICO~\cite{PICO-2L, PICO-60},  XENON1T~\cite{XENON1T}, and so on~\cite{CRESST-II}. In face of this dilemma, several solutions have been proposed in the literature, with the isospin-violating couplings~\cite{Kurylov:2003ra,Giuliani:2005my,Feng:2011vu,Cirigliano:2013zta, Savage:2008er,IVDM1,Frandsen:2011cg, Cline:2011zr, Gao:2011ka, Chen:2014noa}, exothermic scatterings~\cite{Batell:2009vb,Graham:2010ca,Fox:2013pia,McCullough:2013jma,Frandsen:2014ima,Chen:2014tka,Gelmini:2014psa}, and the introduction of a light WIMP-nucleus mediator~\cite{Li:2014vza,Yang:2016wrl} as the three leading mechanisms. After the release of the constraints by LUX~\cite{LUX2013,LUX2015}, SuperCDMS~\cite{SuperCDMS_Ge}, and CDMSlite~\cite{Agnese:2013jaa, Agnese:2015nto}, it was found that only combinations of the above three mechanisms~\cite{DelNobile:2013gba,Gelmini:2014psa,Cirigliano:2013zta,Gresham:2013mua, Chen:2014tka,Witte:2017qsy,Geng:2016uqt}, rather than any single one, could totally eliminate or partially relax the tension between the CDMS-Si excesses and other exclusion limits. In particular, we proposed a unified framework in Ref.~\cite{Geng:2016uqt} by incorporating all of these three mechanisms. Furthermore, with the datasets available in 2015~\cite{LUX2015,  Agnese:2015nto,SuperCDMS_Ge}, we showed that several previously promising models were already excluded at 90$\%$ C.L., including the Ge-phobic exthermic WIMP models in Ref.~\cite{Gelmini:2014psa}, the Xe-phobic WIMP model with nuclear interactions mediated by a light particle in Ref.~\cite{Li:2014vza}, and the isospin-conserving exothermic WIMP models with a light mediator. The only remaining models are those with Xe-phobic exothermic WIMP-nucleus collisions~\cite{Chen:2014tka} with/without a light mediator. Besides the above three mechanisms, there are other proposals based on the effective operator analysis~\cite{Chang:2009yt, Fan:2010gt, Fitzpatrick:2012ix, Fitzpatrick:2012ib, Anand:2013yka, Gresham:2013mua, Catena:2016hoj} and the double disk DM model~\cite{McCullough:2013jma,DDDM1,DDDM2}. %Similar results can be obtained by using the halo-independent data analysis methods in Ref..

Note that it is assumed in Ref.~\cite{Geng:2016uqt} that the WIMP-nucleus interaction is of the standard form for the usual spin-independent (SI) DM direct detection. From the viewpoint of the effective field theory, we only take into account the type-I operators defined in Ref.~\cite{Li:2014vza}.\footnote{The WIMP models based on type-I operators have recently been revisited in \cite{Witte:2017qsy} with the new LUX2016, PandaX-II and PICO-60 dataset.} However, there are also type-II and III generalized SI WIMP-nucleon effective operators~\cite{Li:2014vza}, which are, in particular, characterized by the non-trivial dependences %of the WIMP-nucleus scattering cross sections 
on the momentum transfer or DM velocity. Therefore, it is expected that they can lead to important modifications of the resultant nuclear recoil energy spectra and thus the final fittings. The present paper will explore such effects by focusing on viablity of the above three mechanisms to reconcile the CDMS-Si signals with other upper limits, which completes our study of DM direct detections based on the generalized effective operators. Furthermore, with the latest data from LUX2016~\cite{LUX2016}, PandaX-II~\cite{PandaX16b},  XENON1T~\cite{XENON1T}, and PICO-60~\cite{PICO-60}, it is also timely and necessary to update the current status of the constraints on the WIMP interpretations of CDMS-Si excesses.

The paper is organized as follows. In Sec.~\ref{Sec_Model}, we introduce the Type-II/III generalized effective operators studied in the present paper, and the corresponding observables in the DM direct detections. We summarize our analysis methods for LUX2016, PandaX-II,  XENON1T and PICO-60 in Sec.~\ref{Sec_Fit}. In Sec.~\ref{Sec_Res}, we show our results for various combined mechanisms and two types of operators. Finally, we conclude in Sec.~\ref{Sec_Conc}. 

\section{General Framework for Dark Matter Direct Detection}\label{Sec_Model}
In this section, we briefly summarize our framework to calculate the signals in the DM direct detection, which is the extension of discussion of Type-I effective operators in Ref.~\cite{Geng:2016uqt} to the Type-II/III ones. 

In our benchmark model, it is assumed that there are two kinds of Majorana fermionic WIMP particles, $\chi_H$ and $\chi_L$, comprising  the total DM with an equal density in our Universe. The mass difference between them is small and denoted as $\delta \ll m_{H,L}$. The DM direct detection experiments try to measure the nuclear recoils caused by the WIMP collisions in the detector. At the nucleon level, such a process can be effectively described by the following effective operators~\cite{Li:2014vza}:
\begin{itemize}
\item Type-II Operators:
\begin{eqnarray}\label{O2}
{\cal O}_2 &=& \frac{c_N}{q^2+m_\phi^2}(\bar{\chi}_H \gamma^5 \chi_L + \bar{\chi}_L \gamma^5 \chi_H)(\bar{N}N)\,, \nonumber\\
{\cal O}_{10} &=& \frac{c_N}{q^2+m_\phi^2}(\bar{\chi}_H \sigma^{\mu\nu} \gamma^5 \chi_L + \bar{\chi}_L \sigma^{\mu\nu} \gamma^5 \chi_H)(\bar{N}N)\,,
\end{eqnarray}
\item Type-III Operators:
\begin{eqnarray}\label{O6}
{\cal O}_6 &=& \frac{c_N}{q^2+m_\phi^2} (\bar{\chi}_H \gamma^\mu \gamma^5 \chi_L + \bar{\chi}_L \gamma^\mu \gamma^5 \chi_H) (\bar{N} \gamma_\mu N)\,,
\end{eqnarray}
\end{itemize}  
where $N=(p,n)$ denotes a target nucleon and $c_N$ the corresponding Wilson coefficients. In order to incorporate the light mediator effect, the prefactor before each operator is taken to be of the form of a mediator propagator, where $q=|{\mathbf q}|$ is the magnitude of the 3-momentum transfer and $m_\phi$ the mediator mass. Note that the two operators categorized in the Type II can give rise to the same non-relativistic effective operator, so that they will lead to the same physics in the DM direct detection. Furthermore, with the two WIMP particles, we generically expect that there are two kinds of scattering processes: up- and down-scatterings. However, in order to overcome the mass gap, the up-scatterings can only occur for WIMPs with large enough velocities, which is rare due to the DM velocity distribution in our Galaxy. Therefore, the WIMP-nucleon scatterings off a target nuclide, $T$, are dominated by the exothermic interactions:
\begin{eqnarray}
\chi_H (p_1) + T(p_2) \to \chi_L(p_3) + T(p_4)\,.
\end{eqnarray}
The differential cross section of the WIMP-nucleon scattering can be expressed as follows~\cite{Li:2014vza}:
\begin{eqnarray}\label{XectionN}
\frac{d\sigma_N}{d q^2} (q^2, v) = \frac{\bar{\sigma}_N}{4\mu_{\chi N}^2 v^2} G(q^2, v)\,,
\end{eqnarray}
where $\mu_{\chi N}= m_\chi m_N/(m_\chi + m_N)$ is the reduced mass of the WIMP-nucleon system, and $\bar{\sigma}_N$ is the cross section defined at the reference velocity $v_{\rm ref}\sim 200$~km/s. Due to the lightness of the mediator as well as the WIMP currents in the Type-II/III operators, there are non-trivial dependences of the WIMP-nucleon scatterings on the momentum transfer and WIMP velocity, which is encoded by the factor $G(q^2, v)$ defined as
\begin{eqnarray}
G(q^2, v) = \frac{(q^2_{\rm ref}-q^2_{\rm min})|{\cal M}_{\chi N}(q^2,v)|^2}{\int^{q_{\rm ref}^2}_{q^2_{\rm min}} dq^2 |{\cal M}_{\chi N}(q^2, v_{\rm ref})|^2}\,,
\end{eqnarray}
where $|{\cal M}_{\chi N}|^2$ is the squared matrix element averaged over initial states,  $q^2_{\rm ref}\equiv 4 \mu_{\chi N}^2 v_{\rm ref}^2$ is the reference momentum transfer at which $\bar{\sigma}_{N}$ is defined, and $q^2_{\rm min}$ is related to the energy thresholds of the experiments. Concretely, the corresponding formulae of $G(q^2, v)$ can be explicitly obtained for each type of the operators~\cite{Li:2014vza}:
\begin{itemize}
\item Type-II: 
\begin{equation}
G_{2}(q^2) = \frac{q^2/m_\phi^2}{I_2(q_{\rm min}^2/m_\phi^2, q_{\rm ref}^2/m_\phi^2)(1+q^2/m_\phi^2)^2}\,,
\end{equation} 
where
\begin{eqnarray}
I_2(a,b) = \frac{1}{b-a}\ln\left(\frac{1+b}{1+a}\right)-\frac{1}{(1+a)(1+b)}\,,
%I_2(\frac{q_{\rm min}^2}{m_\phi^2}, \frac{q_{\rm ref}^2}{m_\phi^2}) = \frac{m_\phi^2}{q_{\rm min}^2 - q_{\rm ref}^2}\ln\frac{m_\phi^2+q^2_{\rm ref}}{m_\phi^2 + q_{\rm ref}^2} - \frac{1}{(1+q_{\rm min}^2/m_\phi^2)(1+q_{\rm ref}^2/m_\phi^2)}\,.
\end{eqnarray}

\item Type-III:
\begin{eqnarray}
G_{3} (q^2,v) = \frac{v_{\perp}^2/v_{\rm ref}^2}{I_3(q_{\rm min}^2/m_\phi^2, q_{\rm ref}^2/m_\phi^2)(1+q^2/m_\phi^2)^2}\,,
\end{eqnarray}
where
\begin{eqnarray}
I_3(a,b) = \frac{1}{b(1+a)} - \frac{1}{b(b-a)}\ln\left(\frac{1+b}{1+a}\right) \,,
\end{eqnarray}
where ${\mathbf v}_{\perp} ={\mathbf v} + {\mathbf q}/{(2\mu_{\chi N})} $ is the transverse velocity of DM particle so that $v_\perp^2 = v^2 - q^2/{(4\mu_{\chi N}^2)}$.
\end{itemize}

At the nucleus level, the SI differential cross sections can be transformed into
\begin{eqnarray}\label{XectionT}
\frac{d\sigma_T}{dq^2} = \frac{\bar{\sigma}_p}{4\mu_{\chi p}^2 v^2} [Z+ \xi(A-Z)]^2 G^T(q^2, v)F_T^2(q^2)\,,
\end{eqnarray}
where $A$ and $Z$ are the atomic mass and atomic numbers of the target nucleus. $\xi \equiv c_n/c_p$ is the WIMP coupling ratio between the neutron and proton, which represents the isospin violation effect. $F_T(q^2)$ is the nuclear form factor, which is usually taken to be of the Helm form~\cite{FormFactor} with the same parameters as in Ref.~\cite{Geng:2016uqt}, and  $G^T(q^2,v)$ 
%in Eq.~(\ref{XectionT}) 
encodes the additional DM velocity and/or transfer momentum dependence. Specifically, for Type-II operators, $G^T_2(q^2, v) = G_2(q^2,v)$, while for the Type-III operator, due to an extra factor coming from the nucleon velocity operator ${\bf v}$ acting on the nucleus wave function~\cite{Fitzpatrick:2012ix,Li:2014vza}, $G^T_3$ takes the following form:
\begin{eqnarray}
G^T_3 = \frac{(v^2-q^2/4\mu_{\chi T}^2)/v^2_{\rm ref}}{I_3(q^2_{\rm min}/m_\phi^2,q^2_{\rm ref}/m_\phi^2)(1+q^2/m_\phi^2)^2}\,,
\end{eqnarray}
where $\mu_{\chi T}$ is the WIMP-nucleus reduced mass. 

With the WIMP-nucleus differential cross section in Eq.~(\ref{XectionT}), we can easily obtain the differential recoil event rate per unit detector mass for a nuclide $T$
\begin{eqnarray}
\frac{dR_T}{dE_{\rm nr}} = \frac{\rho_\chi}{m_\chi} \int_{|{\mathbf v}|>v_{\rm min}} d^3 {\mathbf v} v f({\mathbf v}) \frac{d\sigma_T}{dq^2}\,
\end{eqnarray}
where $\rho_\chi = 0.3$~GeV/${\rm cm}^3$ is the local DM energy density, $E_{\rm nr}$ denotes the nuclear recoil energy which can be related to the momentum transfer via $E_{\rm nr}= q^2/(2m_T)$, and $f({\mathbf v})$ is the Galactic DM velocity distribution in the lab frame, which is assumed to be of the Standard Halo Model (SHM)~\cite{Freese:1987wu, Savage:2008er} with the astrophysical parameters taken as in Ref.~\cite{Geng:2016uqt}. The required minimum velocity $v_{\rm min}$ for the integration can be determined kinematically for a nuclear recoil with the energy $E_{\rm nr}$ as follows
\begin{eqnarray}
v_{\rm min} = \frac{1}{\sqrt{2 E_{\rm nr} m_T}} \left|\delta + \frac{m_T E_{\rm nr}}{\mu_{\chi T}}\right|\,.
\end{eqnarray}
For elastic scatterings, $\delta=0$, while for exothermic WIMP scatterings $\delta = m_L-m_H<0$ following the usual conventions~\cite{Graham:2010ca,Frandsen:2014ima,Fox:2013pia}. For the DM velocity integral, we need to specify the operator types:
\begin{itemize}
\item Type-II: 
Since the factor $G^T_2(q^2)$ is only the function of $q^2$, the WIMP velocity distribution integration can be extracted to be
\begin{eqnarray}
\eta(E_{\rm nr},t) = \int_{|{\mathbf v}|>v_{\rm min}} d^3 {\mathbf v} \frac{f({\mathbf v})}{v}\,.
\end{eqnarray}
For the SMH, the above integration can be simplified to be~\cite{Savage:2008er,Li:2014vza}
\begin{eqnarray}\label{eta1}
\eta (E_{\rm nr},t) =\frac{1}{2 N_{\rm esc} v_e} \left[{\rm erf} \left( \frac{v_+}{v_0} \right) -{\rm erf} \left( \frac{v_-}{v_0} \right) - \frac{2}{\sqrt{\pi}}\left(\frac{v_+ - v_-}{v_0}\right) e^{-v^2_{\rm esc}/v_0^2} \right]\, 
\end{eqnarray}
where $v_{\pm} \equiv {\rm min}(v_{\rm min}\pm v_e, v_{\rm esc})$,  $v_e$,$v_{\rm esc}$ and $v_0$ denote the earth, Galactic escape and mean WIMP velocity, respectively, and $N_{\rm esc}$ is the normalization factor, which is given by
\begin{eqnarray}
N_{\rm esc} = {\rm erf}(z) - 2z{\rm exp}(-z^2)/\pi^{1/2}\,,
\end{eqnarray}    
with $z\equiv v_{\rm esc}/v_0$. Note that we take the values of these astrophysical parameters as in Ref.~\cite{Geng:2016uqt}. Therefore, the differential recoil event rate for a target $T$ can be written as
\begin{eqnarray}
\frac{dR_T}{dE_{\rm nr}} = \frac{\rho_\chi}{2m_\chi \mu^2_{\chi T}} \bar{\sigma}_p [Z+\xi(A-Z)]^2 G^T_2(E_{\rm nr}) F_T^2(E_{\rm nr}) \eta(E_{\rm nr},t)\,,
\end{eqnarray} 

\item Type-III: In this case, due to the extra velocity dependence in $G^T_3(q^2,v)$, the differential recoil rate should be written as~\cite{Li:2014vza}:
\begin{eqnarray}
\frac{dR_T}{dE_{\rm nr}} = \frac{\rho_\chi}{2m_\chi \mu^2_{\chi N}} \bar{\sigma}_p [Z+\xi(A-Z)]^2  F_T^2(E_{\rm nr}) (G^{T \prime}_3(E_{\rm nr}) \eta^\prime(E_{\rm nr},t) - G^{T\prime\prime}_3 \eta(E_{\rm nr},t))\,,
\end{eqnarray}
where 
\begin{eqnarray}
G^{T\prime}_3 (E_{\rm nr}) &=& \frac{1/v^2_{\rm ref}}{I_3(q^2_{\rm min}/q_\phi^2, q^2_{\rm ref}/m_\phi^2)(1+q^2/m_\phi^2)^2}\,,\nonumber\\
G^{T\prime\prime}_3 (E_{\rm nr}) &=& \frac{q^2/(4\mu^2_{\chi T}v^2_{\rm ref})}{I_3(q^2_{\rm min}/q_\phi^2, q^2_{\rm ref}/m_\phi^2)(1+q^2/m_\phi^2)^2}\,,\nonumber\\
\eta^\prime (E_{\rm nr},t) &=& \int_{|{\mathbf v}|>v_{\rm min}} d^3{\mathbf v} v f({\mathbf v})\,,\nonumber\\
&=& \frac{v_0}{\sqrt{\pi}N_{\rm esc}} \left[\left( \frac{v_-}{2v_e}+1 \right)e^{-v_-^2/v_0^2}- \left( \frac{v_+}{2v_e}+1 \right)e^{-v_+^2/v_0^2}\right]\nonumber\\
&& + \frac{v_0^2}{4v_e N_{\rm esc}} \left(1+ \frac{2v_e^2}{v_0^2}\right)\left[{\rm erf}\left(\frac{v_+}{v_0}\right)-{\rm erf}\left(\frac{v_-}{v_0}\right)\right] \\
&& -\frac{v_0}{\sqrt{\pi}N_{\rm esc}} \left[2+\frac{1}{3v_e v_0^2}\Big((v_{\rm min}+v_{\rm esc}-v_-)^3-(v_{\rm min}+v_{\rm esc}-v_+)^3\Big)\right]e^{-v_{\rm esc}^2/v_0^2}\,,\nonumber
\end{eqnarray}
where $v_\pm$ are defined as Eq.~(\ref{eta1}).

\end{itemize}

It is impossible for an experiment to measure the nuclear recoil energy with perfect precision. Instead, one measures a proxy, $s$, such as the electron equivalent energy $E_{\rm ee}$ , the prompt scintillation $S1$, the ionization signal $S2$, the heat released, and so on. We can relate these signals to the recoil energy via $s=f_s(E_{\rm nr})$~\cite{Geng:2016uqt,Savage:2008er}. As a result, the total recoil rate per unit detector mass can be described in terms of these variables as below:
\begin{eqnarray}
R(t) = \sum_T g_T \int^\infty_0  dE_{\rm nr} \epsilon (s)  \Phi (f_s(E_{\rm nr}),s_1, s_2)  \frac{dR}{dE_{\rm nr}} \,,
\end{eqnarray}
where the summation is over all target nuclides, each with its mass fraction $g_T$ in the detector material. Here, $\epsilon(s)$ is the detector efficiency, while $\Phi(f_s(E_{\rm nr},s_1, s_2))$ is the response function related to the experimental resolution. If the measured signals are normally distributed about $f_s(E_{\rm nr})$ with a standard deviation $\sigma(s)$, $\Phi(f_s(E_{\rm nr},s_1, s_2))$ is given by:
\begin{eqnarray}
\Phi(f_s(E_{\rm nr}),s_1, s_2) = \frac{1}{2}\left[{\rm erf}\left(\frac{s_2-f_s(E_{\rm nr})}{\sqrt{2}\sigma}\right) - {\rm erf}\left(\frac{s_1-f_s(E_{\rm nr})}{\sqrt{2}\sigma}\right) \right]\,.
\end{eqnarray}

%%%%%%%%%%%%%%%%%%%%%%%%%%%%%%%%%%%%%%%%%%%%%%%%%%%%%%%%%%%%%%%%%%

\section{Experimental Data}\label{Sec_Fit}
In the present paper, we focus on the analysis of compatibility of several DM models to explain the CDMS-Si signals. In particular, we incorporate the latest datasets from XENON100~\cite{XENON100b}, LUX2016~\cite{LUX2016}, PandaX-II~\cite{PandaX16b}, and PICO-60~\cite{PICO-60}, for which the fitting methods are listed below. For other null experiments such as SuperCDMS~\cite{SuperCDMS_Ge}, CDMSlite~\cite{Agnese:2015nto} and  CDEX~\cite{CDEX14, CDEX16}, we follow the same procedures listed in our previous paper~\cite{Geng:2016uqt}, which would not be repeated here.

\subsection{XENON100}
The XENON100 recently updated their data with the total exposure of 477 days, so we obtain the corresponding exclusion bound with the same method as in Ref.~\cite{XENON100b}. 

\subsection{LUX2016}
The LUX search is an experiment with the dual-phase (liquid-gas) xenon time-projection chamber, which measures the prompt scintillation signal $S1$ and their secondary electroluminescence photons $S2$. The LUX bound in this analysis uses the complete LUX 2013+2016 dataset with the exposure $4.47\times 10^4$~kg-day~\cite{LUX2016}, and the threshold of nuclear recoil energy is assumed to be 1.1 keV. The computation of the total number of expected signal events is performed in a way outlined in Ref.~\cite{Geng:2016uqt}, with the detection efficiency extracted from the black curve in Fig.~2 and the $S1$ fractional resolution from Fig.~5 of Ref.~\cite{LUX2016}. For the $S1$ signal, we employ the gain factor $g_1 = 0.117$ and the light yield as the red curve on the slide 13 of Ref.~\cite{LyLUX16}. We obtain the LUX bound by requiring the WIMP scattering cross section corresponding to generating 3.2 events, which was shown in Ref.~\cite{Witte:2017qsy} to agree with the 90$\%$ C.L. exclusion limit very well.  

\subsection{PandaX-II}
The PandaX-II experimental design is quite similar to LUX. It is another dual-phase xenon experiment located at Jinping Underground Laboratory in China. The collaboration has recently released the combined datasets from Run 8 and Run 9, with a total exposure of $3.3\times 10^4$~kg-days~\cite{PandaX16b}. We compute the expected signal number of PandaX-II as for LUX, with the detection efficiency extracted from the black solid curve in Fig. 2 of Ref.~\cite{PandaX16b}. In our analysis, the upper limit is derived by using the Poisson statistics and assuming that there is no candidate event since the observed events in Ref.~\cite{PandaX16b} are consistent with a leaked electron-recoil background.  More recently, the PandaX-II collaboration has shown their new data from Run 10~\cite{PandaX17} with more data but less background events, which leads to the most stringent DM direct detection upper bounds up to now. We estimate the new PandaX-II limits by rescaling of the PandaX-II 2016 results with the exposure of 54.1~ton-days~\cite{PandaX17}.

 \subsection{XENON1T}
Similar to the LUX and PandaX-II, the XENON1T~\cite{XENON1T} is also a xenon-based experiment. With the exposure of $1042\times 34.2$~kg-days, the collaboration has given the SI upper bound a little stronger than the LUX 2016 results. In the present work, we adopt the nuclear recoil detection efficiency as the black curve in Fig.~1 in Ref.~\cite{XENON1T}, and obtain the XENON1T upper bounds with the Poisson statistics without any observed event.

\subsection{PICO-60}
The PICO-60 experiment~\cite{PICO-60} used superheated bubble chambers with ${\rm C_3 F_8}$ as the target to detect acoustic signals in thermodynamic conditions. With the exposure of 1167 kg-days at a thermodynamic threshold of 3.3 keV, no candidate for single nuclear scatterings was found. In our analysis, we only consider the target of fluorine since this nuclide accounts for roughly $80\%$ in mass and has a lower thermodynamic threshold than carbon as shown in Fig.~4 of Ref.~\cite{PICO-2L}. Following Ref.~\cite{Witte:2017qsy}
, we take the nuclear recoil threshold to be 6~keV corresponding to the 3.2~keV thermodynamic threshold in Fig.~4 of \cite{PICO-2L}. The bound for PICO-60 is obtained by using the Poisson statistics with the assumption of no event observed and no background subtraction. 

\section{Fitting Results}\label{Sec_Res}
In this section, we consider a couple of combinations of the three typical mechanisms, such as isospin-violation, exothermic scattering and a light mediator, trying to make the CDMS-Si data compatible with the other null experiments. In Ref.~\cite{Geng:2016uqt}, we systematically studied constraints on the CDMS-Si signal after the 2015 data of LUX~\cite{LUX2015} and CDMSlite~\cite{Agnese:2015nto} for several typical combinations, including new exothermic DM models scattering via a light mediator with/without the isospin-violating interactions. However, the analysis in Ref.~\cite{Geng:2016uqt} was only performed for Type-I effective operators, so it is necessary to extend such discussions to other two types of operators. Furthermore, the advent of the new datasets from LUX2016~\cite{LUX2016}, PandaX-II~\cite{PandaX16b, PandaX17}, XENON1T~\cite{XENON1T} and PICO-60~\cite{PICO-60} have provided new challenges to above interpretations, which have recently been investigated in Ref.~\cite{Witte:2017qsy} for Type-I effective operators\footnote{Ref.~\cite{Witte:2017qsy} only considered the effects of isospin-violation and exothermic scatterings without a light mediator. But the results there can be applied even to the case with a light mediator, since our previous study in Ref.~\cite{Geng:2016uqt} shows that the light mediator only affects the fitting very mildly. }. Therefore, it is timely and necessary to visit the validity of these mechanisms in the context of the Type-II/III operators under new datasets.

\subsection{Dark Matter Model with Isospin-Conserving Elastic Contact Interactions}
\begin{figure}[ht]
\includegraphics[scale = 0.32, angle=270]{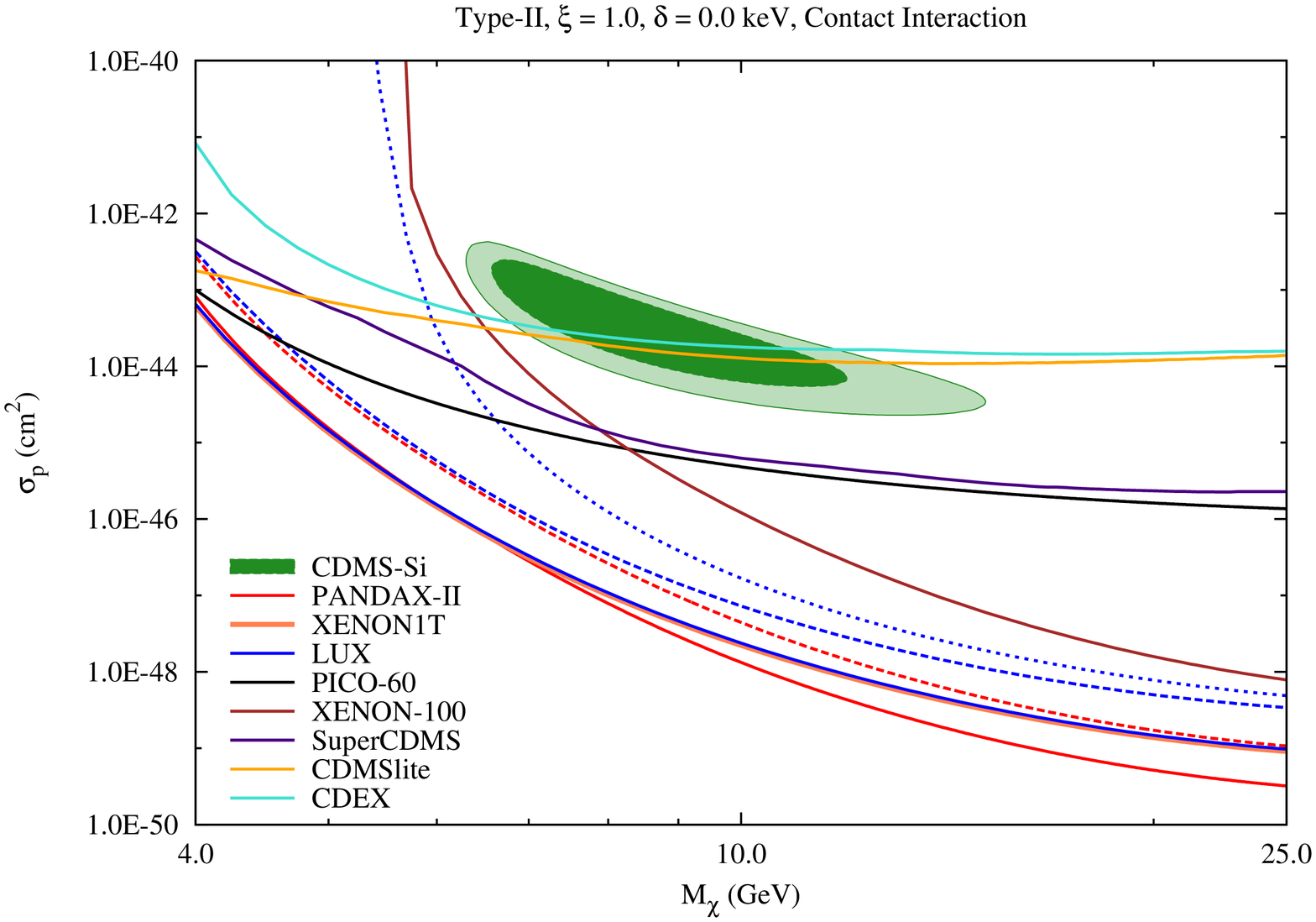}
\includegraphics[scale = 0.32, angle=270]{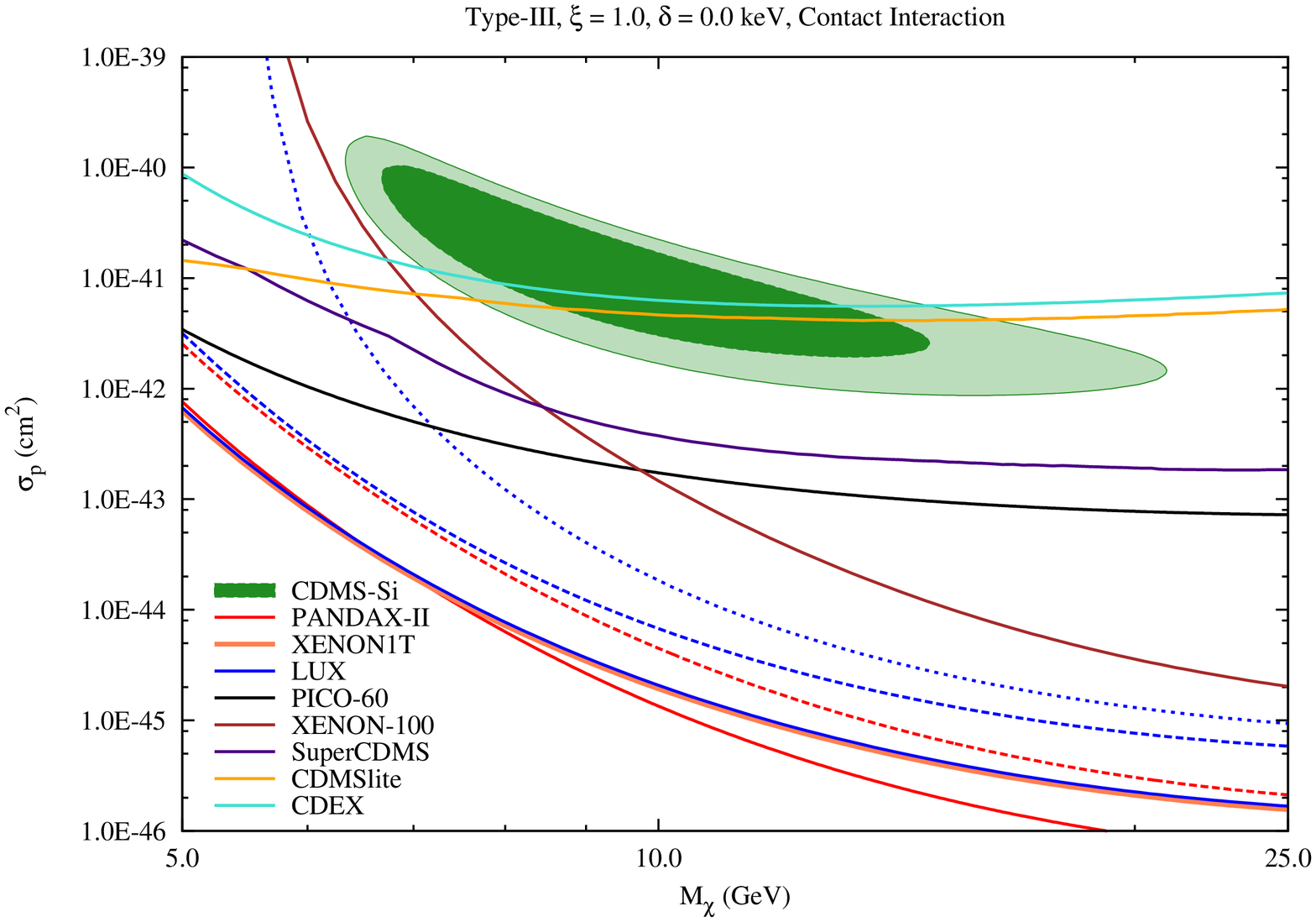}
\caption{CDMS-Si signal regions at 90$\%$ C.L. (light green) and  68$\%$ C.L. (dark green) and exclusion limits at 90$\%$ C.L. of SuperCDMS (purple), CDMSlite (yellow), LUX2013 (blue dotted), LUX2015 (blue dashed), LUX2016 (blue solid),  PandaX-II-2016 (red dashed), PandaX-II-2017 (red solid), XENON1T(pink), XENON100 (brown) and CDEX (cyan) for Type-II (Left Panel) and Type-III (Right Panel) operators with isospin-conserving elastic contact DM nuclear scatterings.}\label{Res_x1d0c}
\end{figure}

Fig.~\ref{Res_x1d0c} shows the fitting results of the DM models with isospin-conserving elastic nuclear scatterings via a contact interaction for both Type-II and Type-III effective operators. It is interesting to note that the main difference in fittings of the two types of operators is overall scales of the WIMP-proton scattering cross sections for CDMS-Si signals and various upper limits, with the Type-III results typically larger than Type-II by about two orders of magnitude, a feature which is very generic as shown below for other DM scenarios. Moreover, the Type-III operator predicts the CDMS-Si signal regions to extend to the higher DM mass range than that of the Type-II operators. Despite these distinctions, it is evident that the whole $90\%$ C.L. CDMS-Si contours in both plots have already been excluded by the null results from LUX2013 and SuperCDMS, needless to say the more stringent ones from LUX2016, PandaX-II,  XENON1T and PICO-60. Note that we choose the mediator mass to be $m_\phi = 200$~MeV here, since it is already heavy enough compared with the typical momentum transfer so that the light mediator effects are effectively turned off.     

\subsection{Exothermic Dark Matter with Isospin-Violating Couplings}
In this subsection, we study several typical isospin-violating exothermic DM models. Two examples in this class include the exothermic Xe-phobic and Ge-phobic WIMPs, each with the isospin parameter to be $\xi = -0.7$ and $\xi = -0.8$, respectively, which are intended to maximally reduce the DM sensitivity to xenon and germanium. For exothermic DM models, we will consider the cases with the WIMP mass splitting to be $\delta = -50$ and $-200$~keV. Note that $|\delta| = 200$~keV was shown to be the largest value allowed by the three observed CDMS-Si events to be WIMP-nucleus scattering candidates simultaneously~\cite{Fox:2013pia, Frandsen:2014ima, Gelmini:2014psa, Witte:2017qsy}. Therefore, in the present paper, we restrict our attention to the models with $|\delta| \leq 200$~keV. 

\begin{figure}[ht]
\includegraphics[scale = 0.32, angle=270]{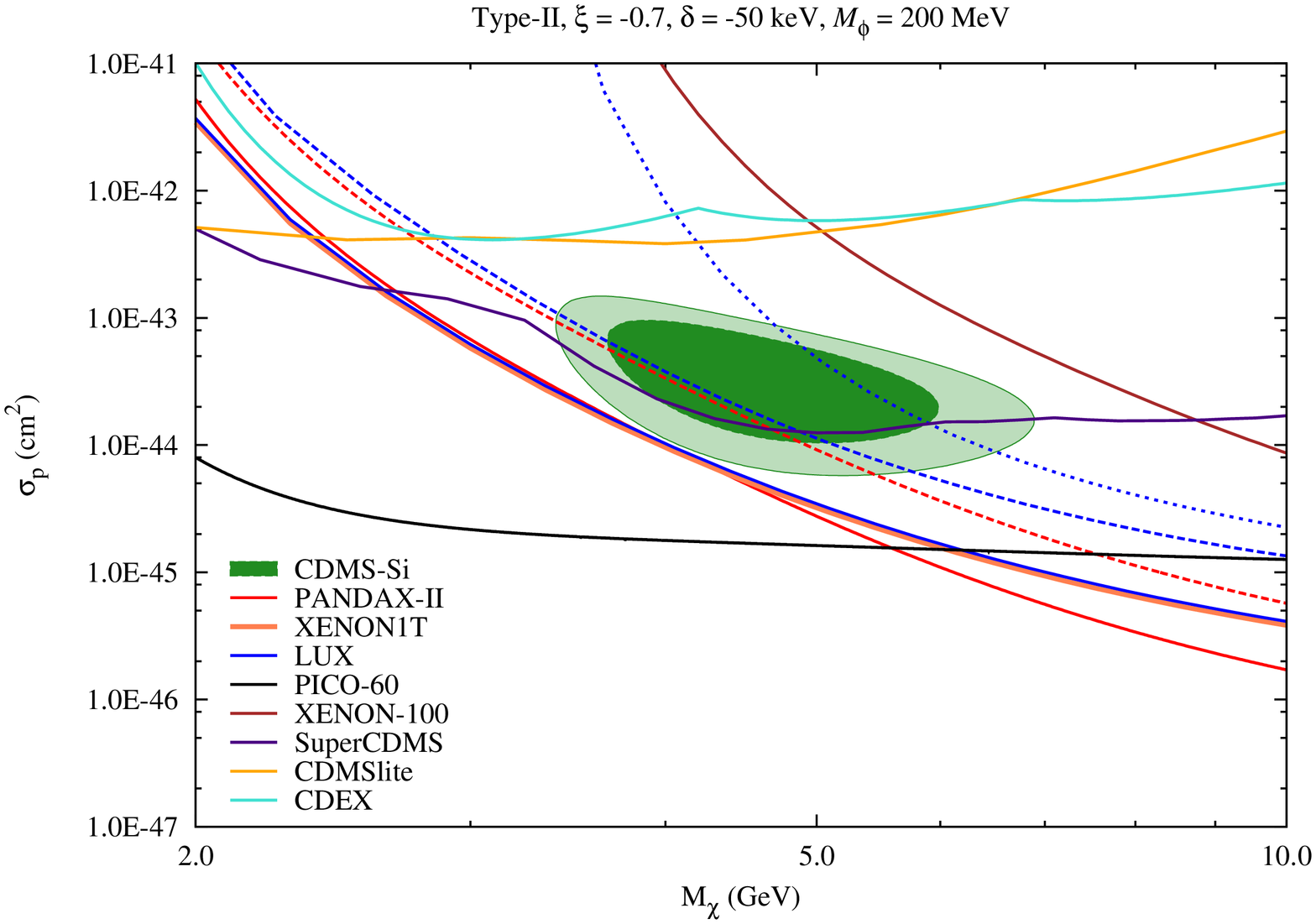}
\includegraphics[scale = 0.32, angle=270]{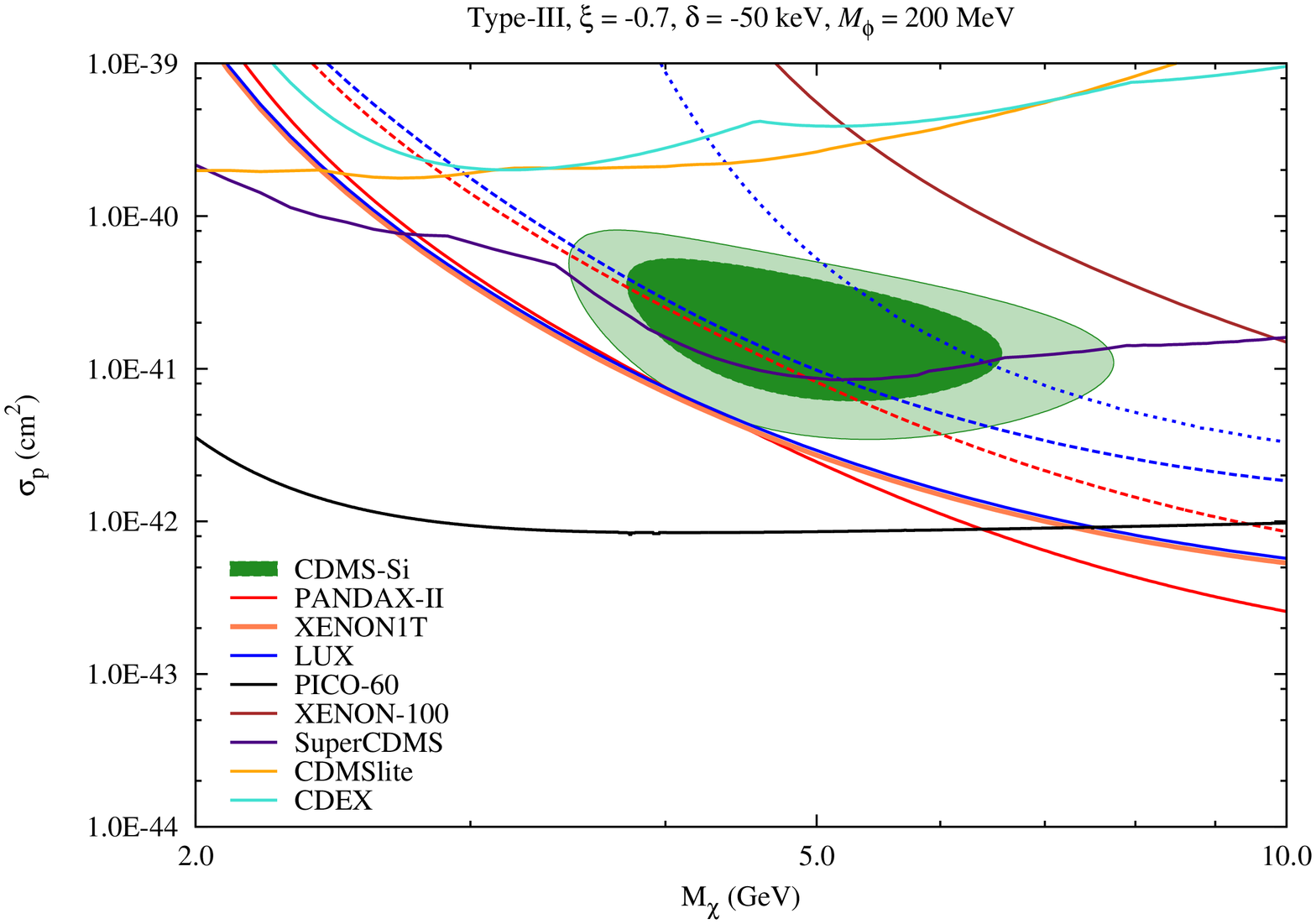}
\caption{Legend is the same as in Fig.~\ref{Res_x1d0c} except for Xe-phobic exthermic WIMP models with $\xi = -0.7$, $\delta = -50$~keV and contact interactions. }\label{Res_x07d50c}
\end{figure}
\begin{figure}[ht]
\includegraphics[scale = 0.32, angle=270]{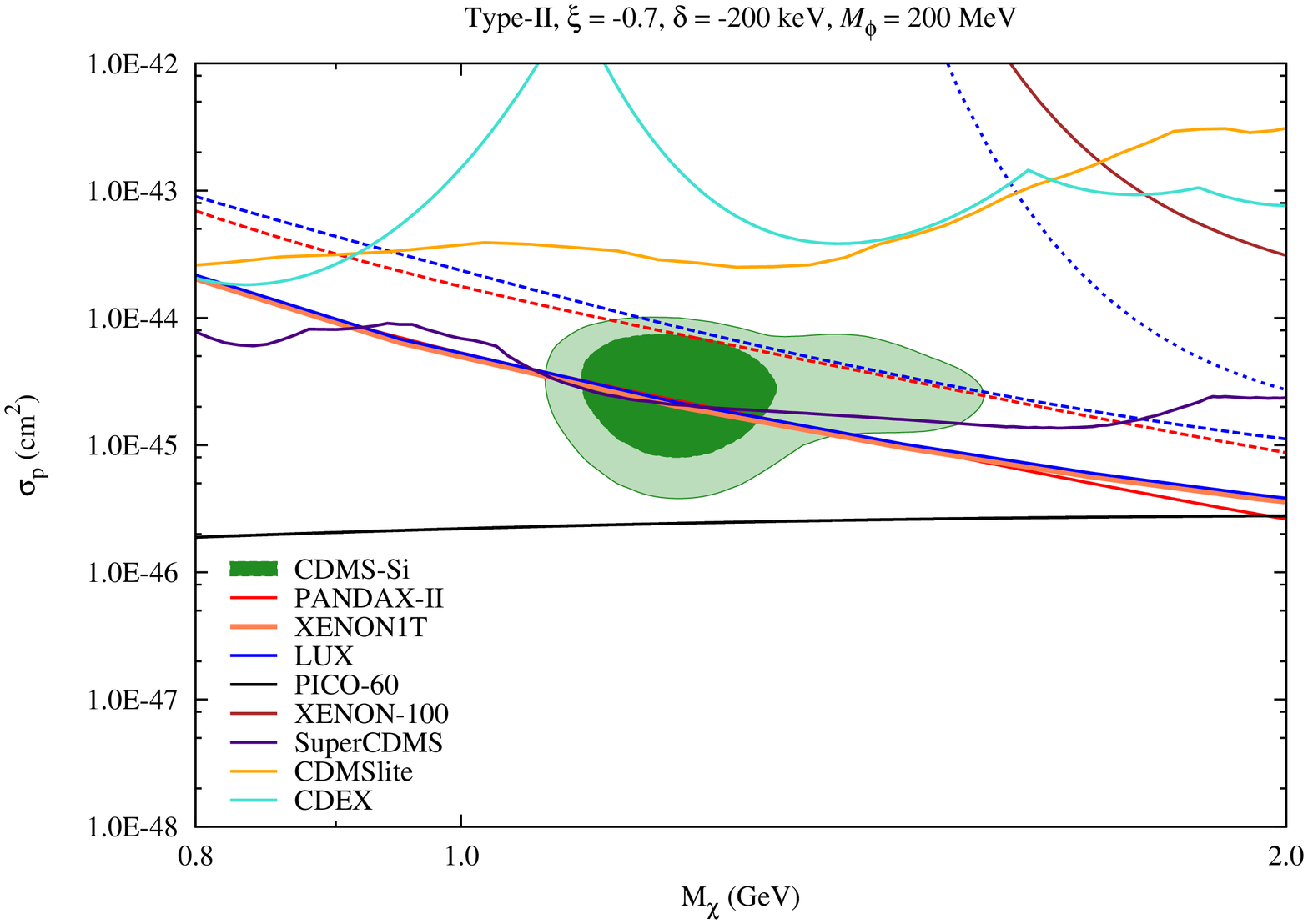}
\includegraphics[scale = 0.32, angle=270]{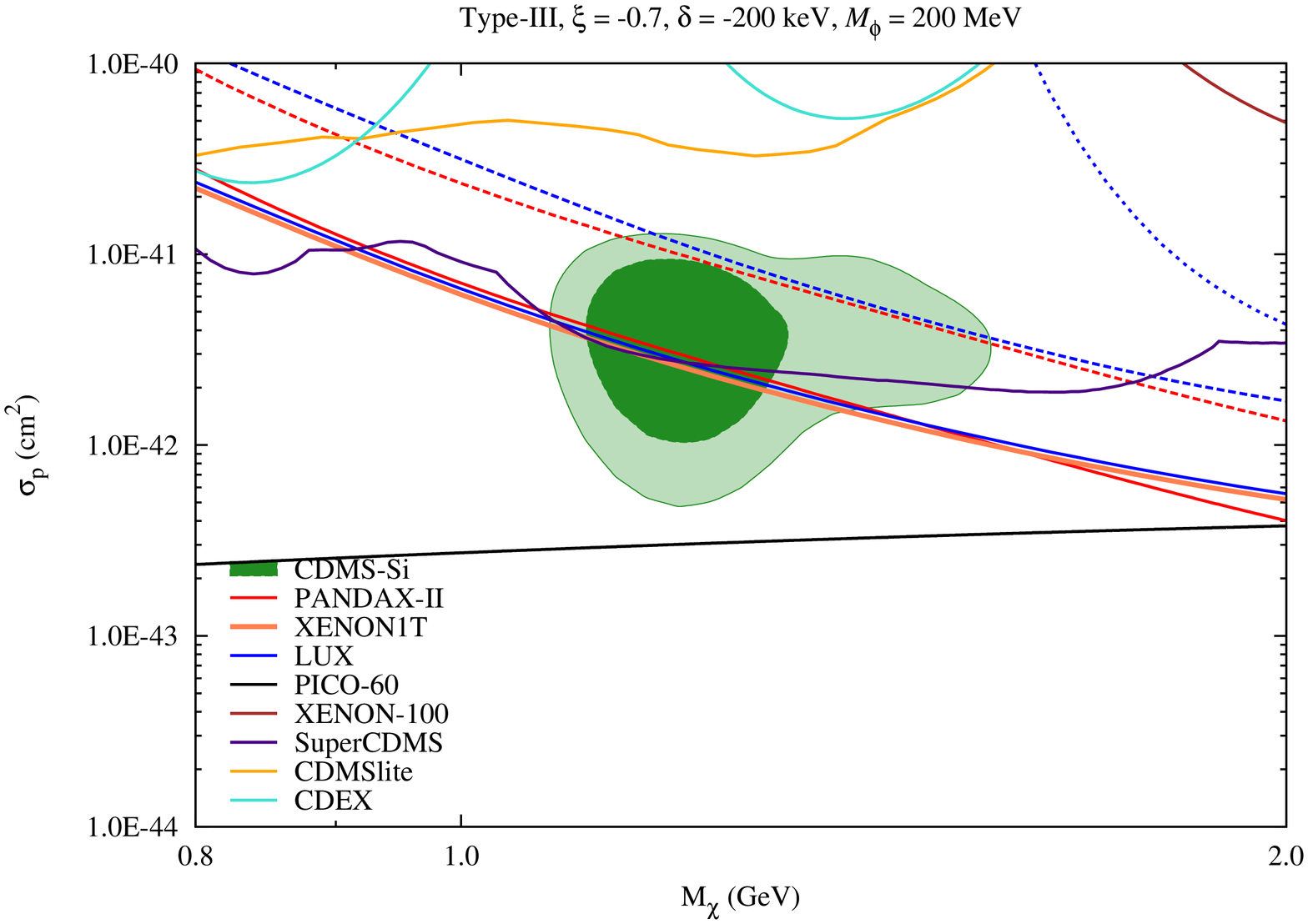}
\caption{Legend is the same as in Fig.~\ref{Res_x1d0c} except for Xe-phobic exthermic WIMP models with $\xi = -0.7$, $\delta = -200$~keV and contact interactions. }\label{Res_x07d200c}
\end{figure}

We present in Figs.~\ref{Res_x07d50c} and \ref{Res_x07d200c} the analyses for Xe-phobic exothermic DM models with a mass difference $\delta= -50$ and $-200$~keV, respectively. From the two panels in each plot, it is clear that both types of WIMP operators give the very similar fitting results to each other, except the overall scales of WIMP-nucleon cross sections. For $\delta=-50$~keV, the data before 2016, like LUX2015 and CDMSlite, could only restrict $68\%$ C.L. CDMS-Si region, and a part of $90\%$ C.L. region was still allowed. But now the whole $90\%$ C.L. CDMS-Si region is excluded by the $90\%$ upper limits from LUX2016, PandaX-II,  XENON1T, and PICO-60. 

When $\delta = -200$~keV, as illustrated in Fig.~\ref{Res_x07d200c}, compared with the constraints imposed by LUX2015, the $68\%$ and $90\%$ C.L. CDMS-Si regions allowed by the data of LUX2016 and PandaX-II are shrunk greatly, which are further rejected by the new PICO-60 $90\%$ C.L. bound.  Note that the inclusion of latest data from XENON1T and PandaX-II-2017, the bounds from xenon-based experiments do not improve significantly, since the xenon detectors lose their sensitivities very quickly for low DM masses which is favored by the CDMS-Si signals.

\begin{figure}[ht]
\includegraphics[scale = 0.32, angle=270]{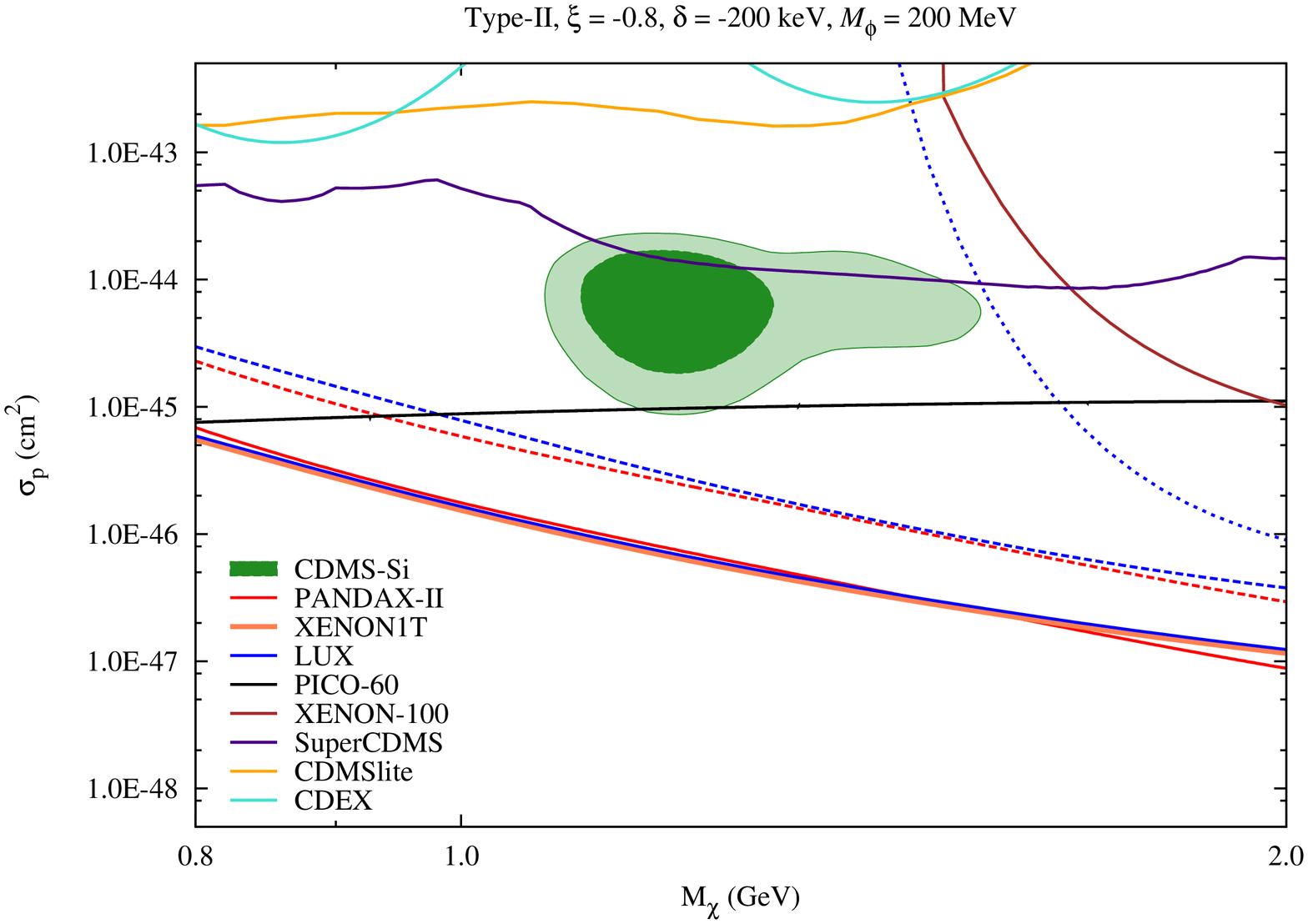}
\includegraphics[scale = 0.32, angle=270]{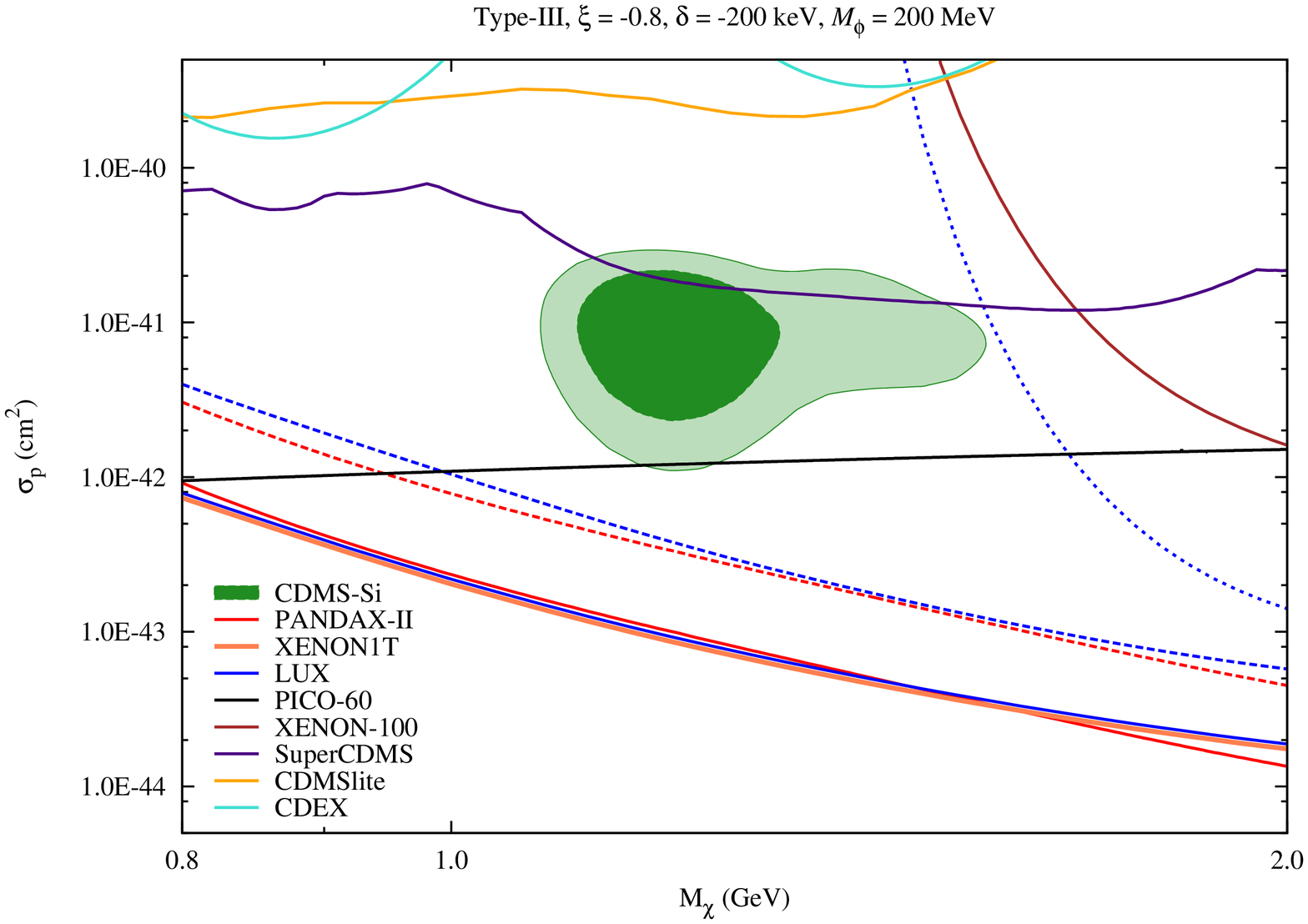}
\caption{Legend is the same as in Fig.~\ref{Res_x1d0c} except for Ge-phobic exthermic WIMP models with $\xi = -0.8$, $\delta = -200$~keV and contact interactions. }\label{Res_x08d200c}
\end{figure}

For the Ge-phobic exothermic DM with a maximal WIMP mass gap $\delta=-200$~keV, it is clear in Fig.~\ref{Res_x08d200c} that for both types of operators, the CDMS-Si $90\%$ C.L. regions are in great tension with the LUX2015/2016 and PandaX-II data.

In summary, the previously promising exothermic DM models with either Xe-phobic or Ge-phobic interactions for both types of operators are now facing a great challenge in reconciling the CDMS-Si signals with the recent experimental results from LUX2016, PandaX-II,  XENON1T and PICO-60.

\subsection{Xe-phobic Dark Matter with a Light Mediator}
\begin{figure}[ht]
\includegraphics[scale = 0.32, angle=270]{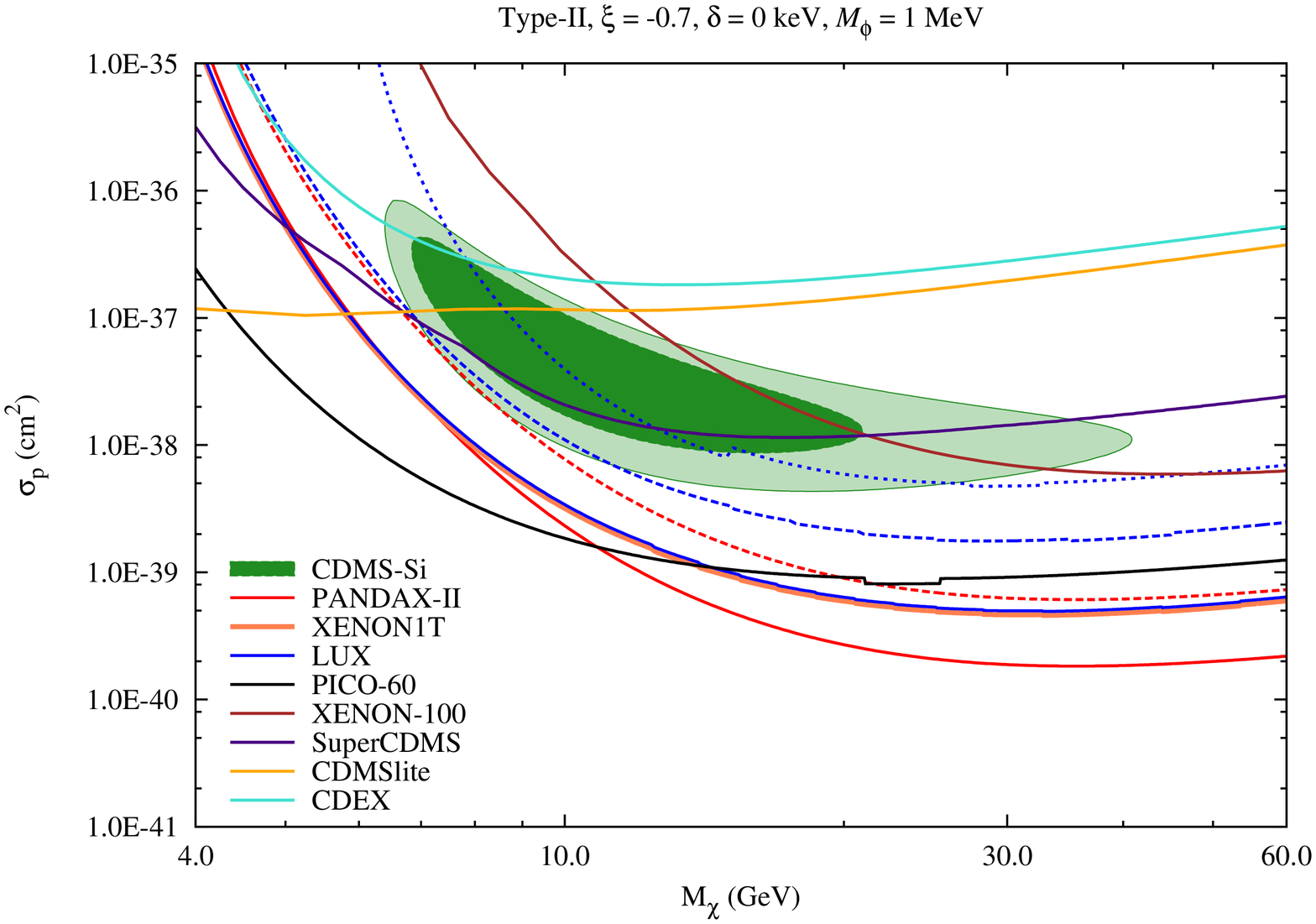}
\includegraphics[scale = 0.32, angle=270]{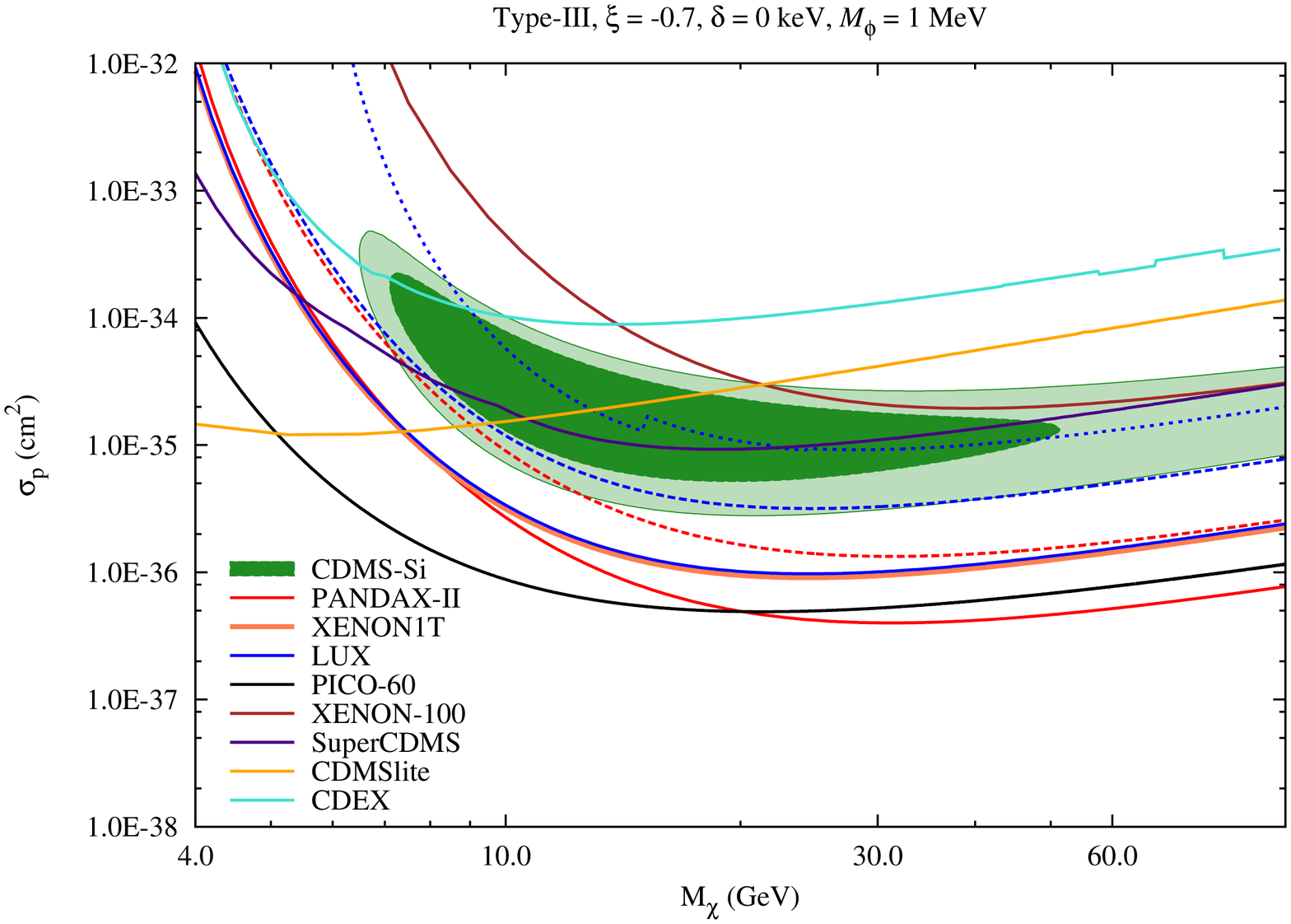}
\caption{Legend is the same as in Fig.~\ref{Res_x1d0c} except for Xe-phobic elastic WIMP models with $\xi = -0.7$, $\delta =  0$~keV and a light mediator of $m_\phi = 1$~MeV. }\label{Res_x07d0m1}
\end{figure}

We now turn to the DM models with a light mediator for the Type-II/III effective operators. Here, we choose the mediator mass to be $m_\phi = 1$~MeV, which is small compared with the typical direct search momentum transfer of ${\cal O}(10)$~MeV so that the light mediator effect would be saturated maximally. Firstly, we consider the Xe-phobic WIMP with elastic scatterings, and the results are shown in Fig.~\ref{Res_x07d0m1}. Note that the distinction of CDMS-Si signal regions for the two types of operators is clear in this case, in which the mass of Type-II WIMPs is determined to be around $6 \sim 40$~GeV, whereas for the Type-III operator the mass can extend to more than 100~GeV. However, for both operator types, the whole $90\%$ signal region of CDMS-Si has been excluded by LUX2016,  PandaX-II and XENON1T upper limits.

\subsection{Exothermic Dark Matter with a Light Mediator}
\begin{figure}[ht]
\includegraphics[scale = 0.32, angle=270]{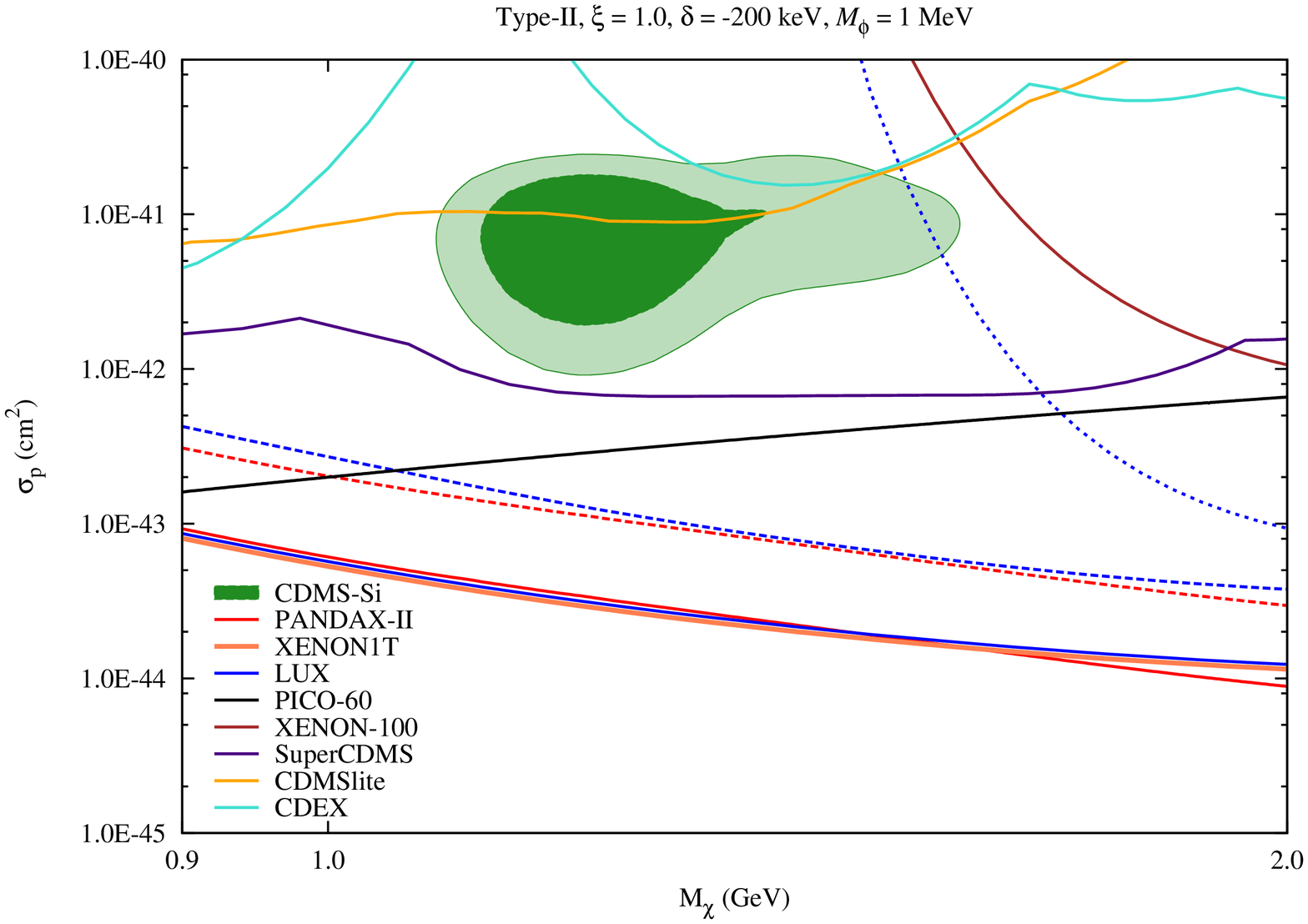}
\includegraphics[scale = 0.32, angle=270]{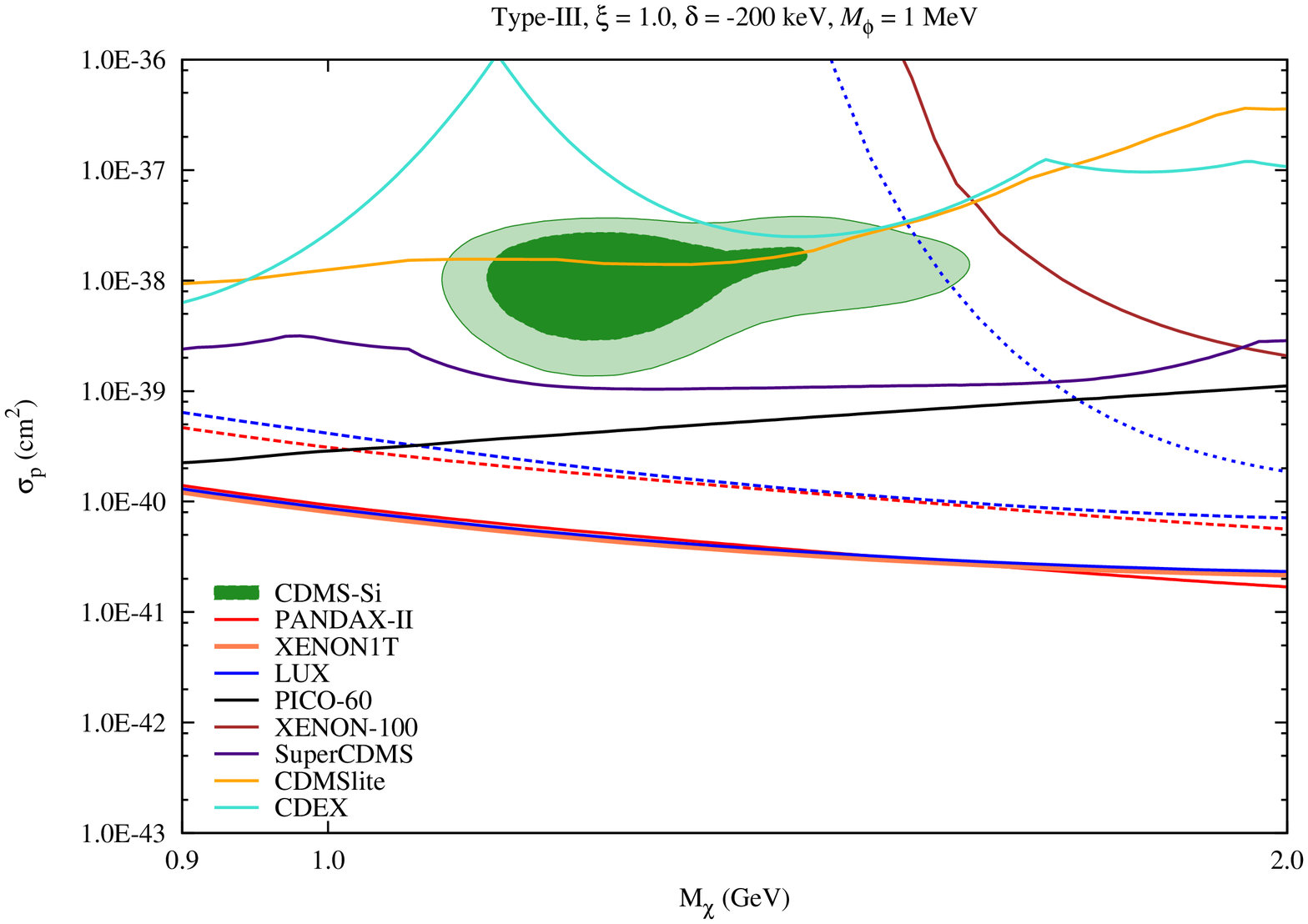}
\caption{Legend is the same as in Fig.~\ref{Res_x1d0c} except for isospin-conserving exothermic WIMP models with $\xi = 1.0$, $\delta =  0$~keV and a light mediator of $m_\phi = 1$~MeV. }\label{Res_x0d200m1}
\end{figure}

We explore the exothermic DM case with a light mediator and isospin-conserving couplings. We adopt the extremal values for both the mediator mass and the WIMP gap, {\it i.e.}, $m_\phi = 1$~MeV and $\delta = -200$~keV, in order to maximize their effects. However, as shown in Fig.~\ref{Res_x0d200m1},  neither types of operators can make the CDMS-Si preferred parameter spaces survive the combined exclusion limits from SuperCDMS, LUX2015/2016, PandaX-II,  XENON1T, and PICO-60. Hence, it can be concluded, without combining isospin-violating couplings, this model is unable to relieve the significant tension.  

\subsection{Exothermic Dark Matter with a Light Mediator and Isospin-Violation}
\begin{figure}[ht]
\includegraphics[scale = 0.32, angle=270]{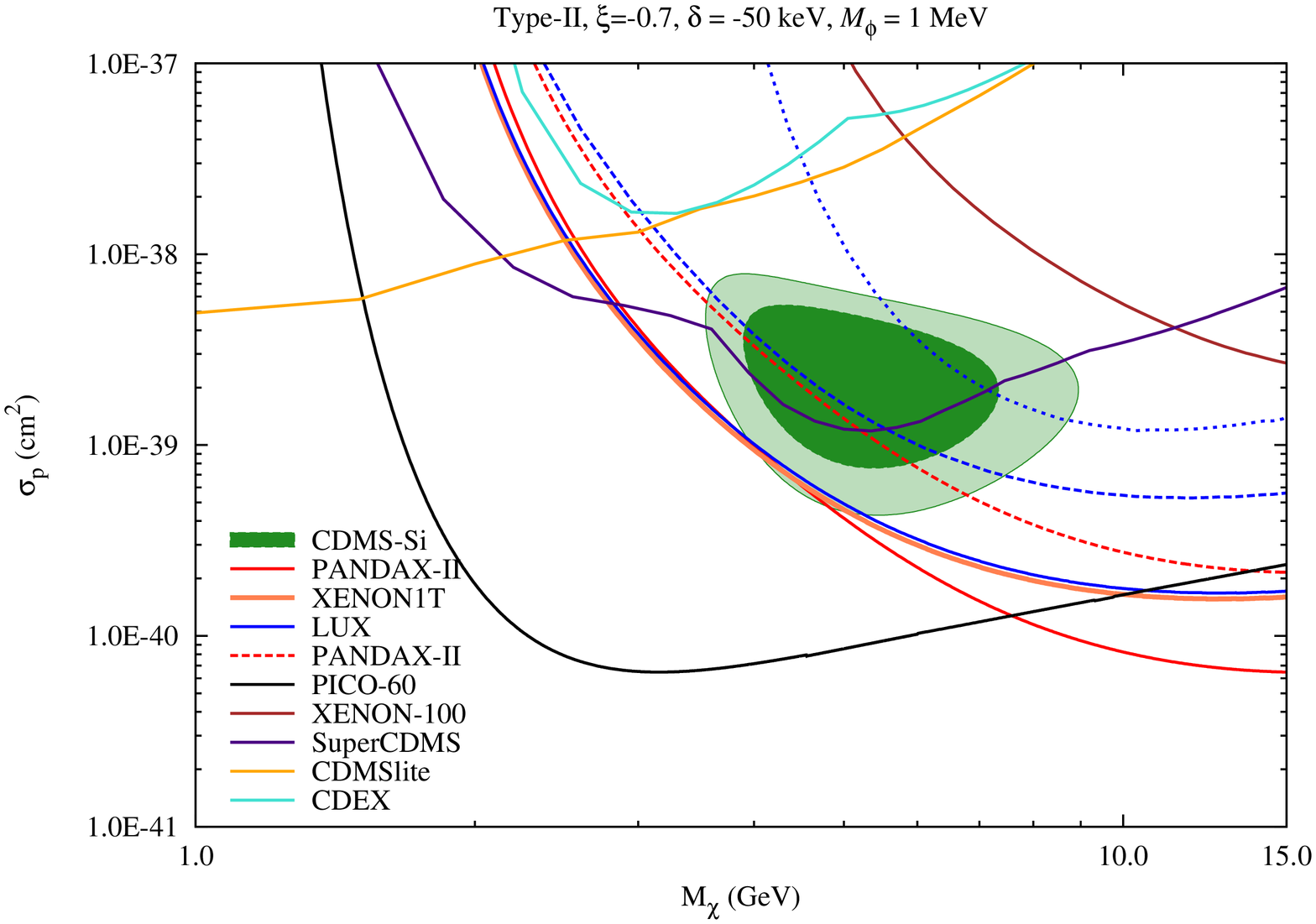}
\includegraphics[scale = 0.32, angle=270]{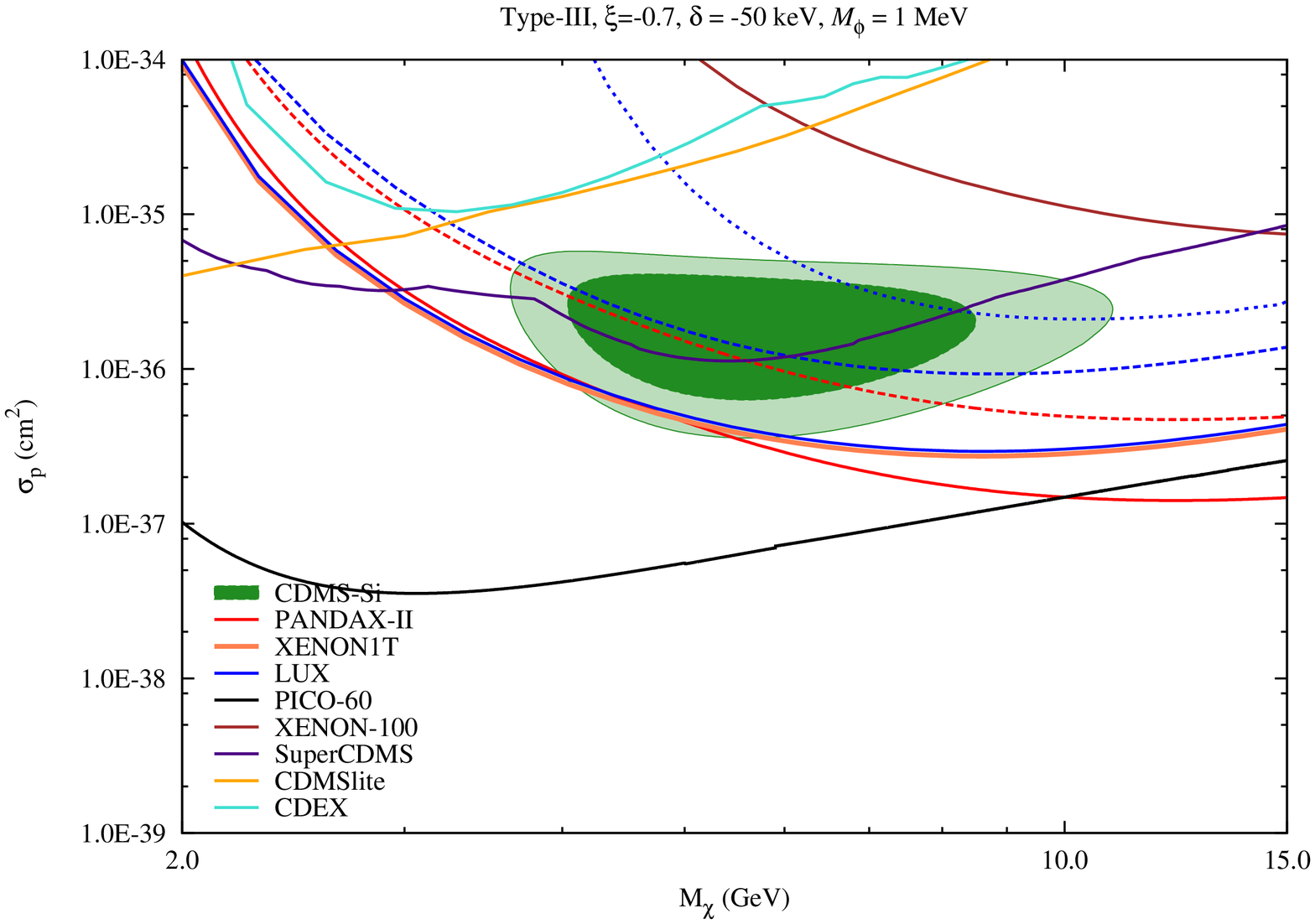}
\caption{Legend is the same as in Fig.~\ref{Res_x1d0c} except for Xe-phobic exothermic WIMP models with $\xi = -0.7$, $\delta =  -50$~keV and a light mediator of $m_\phi = 1$~MeV. }\label{Res_x07d50m1t}
\end{figure}

\begin{figure}[ht]
\includegraphics[scale = 0.32, angle=270]{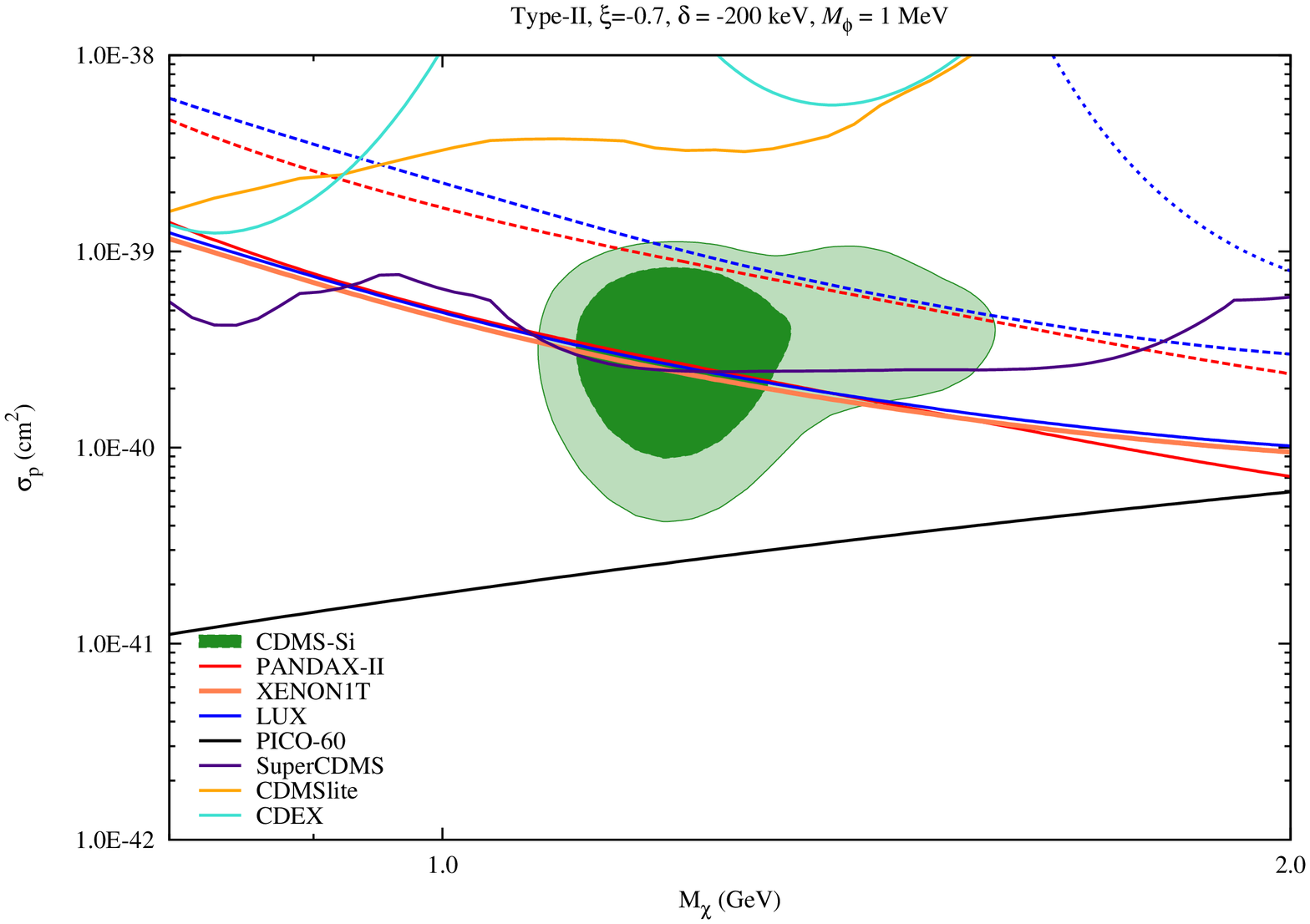}
\includegraphics[scale = 0.32, angle=270]{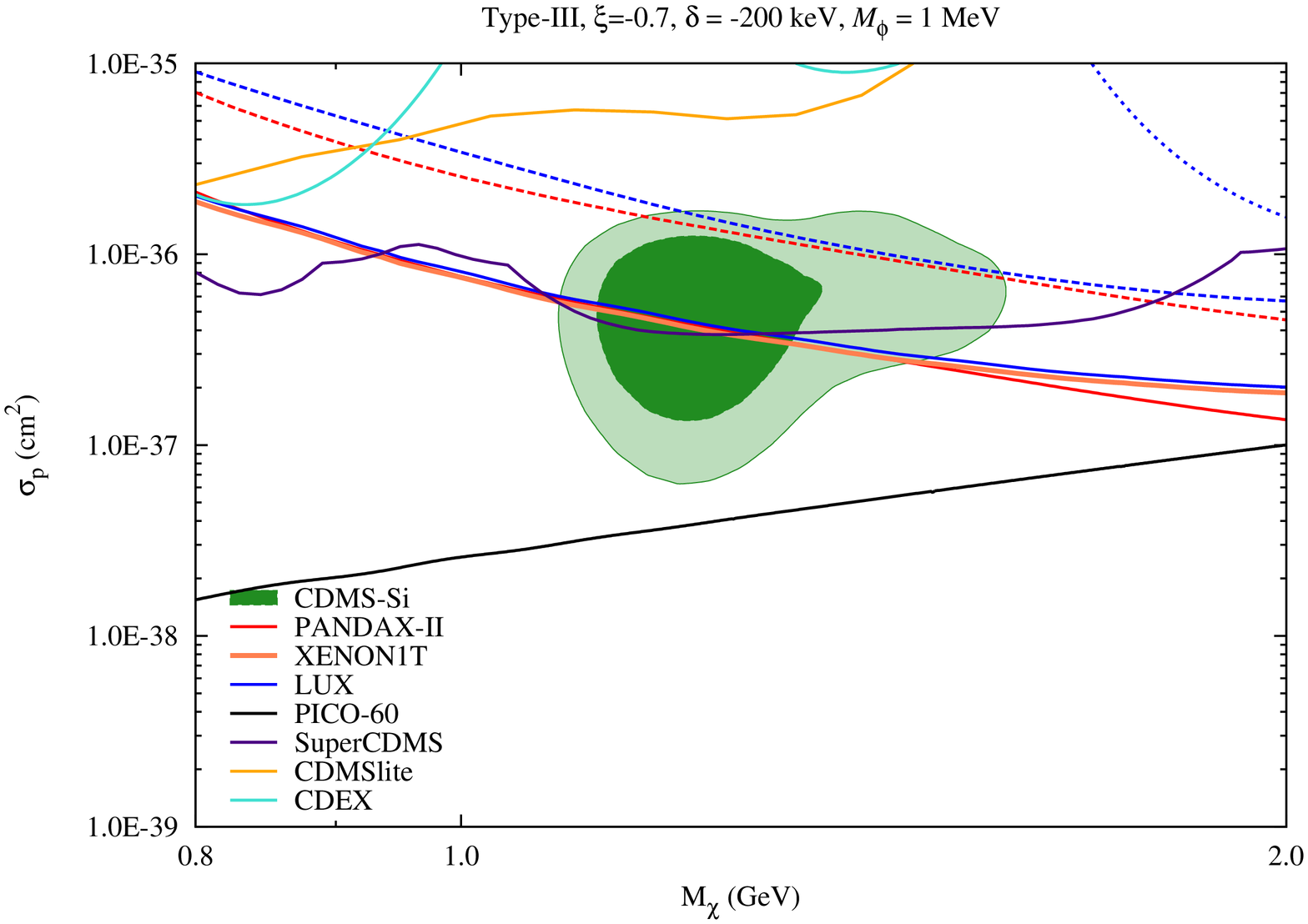}
\caption{Legend is the same as in Fig.~\ref{Res_x1d0c} except for Xe-phobic exothermic WIMP models with $\xi = -0.7$, $\delta =  -200$~keV and a light mediator of $m_\phi = 1$~MeV. }\label{Res_x07d200m1t}
\end{figure}

Finally, we would like to investigate the cases which combine all the three mechanisms. Concretely, the isospin-violating parameter is chosen to be Xe-phobic $\xi=-0.7$, and the mediator is taken to be light with $m_\phi = 1$~MeV, but two typical mass gaps, $\delta = -50$~keV and $-200$~keV, are considered. The fitting results for both operator types are displayed in Figs.~\ref{Res_x07d50m1t} and \ref{Res_x07d200m1t}, which are shown to be very similar to the cases with contact interactions in Figs.~\ref{Res_x07d50c} and \ref{Res_x07d200c}, meaning that the light mediator affects the fittings very mildly compared with the other two effects. 
For $\delta=-50$~keV, the LUX2016, PandaX-II  and XENON1T constraints in Fig.~\ref{Res_x07d50m1t} cannot entirely exclude the $90\%$ C.L. CDMS-Si region, which is, however, definitely excluded by the PICO-60 $90\%$ C.L. bound. When the mass gap increases to $\delta=-200$~keV, nearly half of $68\%$ and $90\%$ C.L. CDMS-Si regions are still allowed by the SuperCDMS,  LUX2016, PandaX-II-2017 and XENON1T $90\%$ C.L. upper limits, but they cannot survive by taking the new PICO-60 constraint into account. 

\section{Conclusions}\label{Sec_Conc}
We have extended our previous work in Ref.~\cite{Geng:2016uqt} on the viability of the SI DM interpretation of CDMS-Si signal to the type-II and type-III generalized effective operators listed in Ref.~\cite{Li:2014vza}, which are characterized by the extra non-trivial dependence of the momentum transfer and DM velocity. We have considered in this framework the various combinations of three typical mechanisms of the isospin violation, exothermic interactions and a light mediator. It is found that the different choices of effective operators cannot improve the situation much. Except for the minor shape variation of the fitted CDMS-Si signal regions and the upper limits, the most notable effects of choosing different operators are the rescaling of overall magnitudes of the predicted CDMS-Si cross sections and various upper limits, but the relative positions do not change greatly. Hence, the models that are excluded with the type-I operator are also strongly disfavored by using the type-II/III operators, including the isospin-conserving and Ge-phobic scenarios.

Moreover, the latest LUX2016, PandaX-II,  XENON1T and PICO-60 results further strengthen the constraints on the previously allowed Xe-phobic exothermic models with/without a light mediator. For models with a mass gap of $\delta=-50$~keV and a light mediator of $m_\phi=1$~MeV, the $68\%$~C.L. CDMS-Si region is totally excluded by LUX2016, PandaX-II and XENON1T, and the $90\%$~C.L. region by PICO-60. For the models with the maximal gap $\delta=-200$~keV, regardless of the nuclear recoils mediated by a light mediator or not, while the  LUX2016, PandaX-II and XENON1T still allow a fraction of the CDMS-Si signal region, PICO-60 rejects it completely. Therefore, we are forced to conclude that it is very implausible to interpret the CDMS-Si anomaly by using the SI DM models with isospin-violation, exothermic scattering and/or a light mediator, no matter what effective operators are realized.

 Finally, one possible concern for exothermic scatterings is that the lifetime of the heavy DM state should be longer than the age of the Universe, which is crucial for the present scenario to work. Let us estimate the lifetimes of heavy DM for both operator types in Eqs.~(\ref{O2}) and (\ref{O6}). For the Type-II operator, we take the operator ${\cal O}_2$ as an example. According to the definition of the reference DM nuclear recoil cross section $\bar{\sigma}_N$ in Eq.~(\ref{XectionN}), we can obtain it for ${\cal O}_2$ as follows:
\begin{eqnarray}
\bar{\sigma}^{{\cal O}_2}_N = \frac{c_N^2 m_N^2}{4\pi (m_N+m_\chi)^2} \left[\frac{1}{q^2_{\rm ref}-q^2_{\rm min}}\ln\frac{m_\phi^2+q_{\rm ref}^2}{m_\phi^2+q_{\rm min}^2}  - \frac{m_\phi^2}{(m_\phi^2+q_{\rm ref}^2) (m_\phi^2+q_{\rm min}^2)} \right]\,.
\end{eqnarray}
If we take $m_\chi = 1.2$~GeV and $\bar{\sigma}_p = 2\times 10^{-40}~{\rm cm}^2$ yielded by fitting CDMS-Si signals for the extreme case with $\xi = -0.7$, $\delta = -200$~keV and $m_{\phi} = 1$~MeV, we can estimate the Wilson coefficient $c_p \approx 1.7\times 10^{-8}$. Note that the fastest decay mode for the heavy DM to the lighter one through the operator ${\cal O}_2$ is $\chi_H \to \chi_L + 2\gamma$, in which the decay rate is given by
\begin{eqnarray}
\Gamma_{{\cal O}_2} = \frac{\alpha^2 c_p^2 \delta^8}{2^8 3^3 \pi^5 m_{\chi} m_p^2 m_\phi^4}\,,
\end{eqnarray}   
where $\alpha$ is the fine structure constant. Thus, with the above typical DM parameters, the lifetime of the heavy DM is $4.87 \times 10^{19}$~s, which is much longer than the age of the Universe $\tau_U = 4.3\times 10^{17}$~s. The calculations with the operator ${\cal O}_{10}$ gives the similar results.

As for the Type-III operator, {\it i.e.} ${\cal O}_6$, the reference nucleon-DM scattering cross section is given by
\begin{eqnarray}
\bar{\sigma}_N^{{\cal O}_6} &=& \frac{c_N^2}{4\pi}\Bigg(\frac{q_{\rm ref}^2}{(m_\phi^2+q_{\rm ref}^2)(m_\phi^2 + q_{\rm min}^2)} \nonumber\\
& & - \left[\frac{1}{q_{\rm ref}^2 - q_{\rm min}^2}\ln\frac{m_\phi^2+q_{\rm ref}^2}{m_\phi^2 + q_{\rm min}^2} - \frac{m_\phi^2}{(m_\phi^2 + q_{\rm ref}^2)(m_\phi^2 + q_{\rm min}^2)}\right]\Bigg)\,.
\end{eqnarray}
By fitting the CDMS-Si dataset, it is shown on the right panel of Fig.~8 that the DM mass is around 1.2~GeV and the nuclear recoil cross section is of ${\cal O}(4\times 10^{-37}~{\rm cm^2})$, so that the Wilson coefficient is estimated to be $c_p = 2.5\times 10^{-7}$. According to Refs.~\cite{Finkbeiner:2009mi,Batell:2009vb}, it is shown that the leading heavy DM decay mode is $\chi_H \to \chi_L + 3\gamma$ for the vector SM current. We adopt the approximated decay rate formula in Ref.~\cite{Batell:2009vb}, given by
\begin{eqnarray}
\Gamma_{{\cal O}_6} \approx \frac{17 \alpha^3 c_p^2}{2^9 3^6 5^3 \pi^4} \frac{\delta^{13}}{m_\phi^4 m_p^8}\,,
\end{eqnarray}   
where the extra factors in the denominator compared with Ref.~\cite{Batell:2009vb} come from our definition of the Wilson coefficient $c_N$, and we have replaced the original electron loop with the corresponding proton loop in the Feynman diagram. This leads to the extremely long heavy DM lifetime $\tau_{\chi_H} \approx 5\times 10^{39}$~s, which is certainly longer than the age of Universe. Therefore, it is concluded that the heavy DM is cosmological stable, no matter what types of operators are considered.

\appendix

%%%%%%%%%%%%%%%%%%%%%%%%%%%%%%%%%%%%%%%%%%%%%%%%%%%%%%%%%%%%%%
%%%%%%%%%%%%%%%%%%%%%%%%%%%%%%%%%%%%%%%%%%%%%%%%%%%%%%%%%%%%%%
%%%%%%%%%%%%%%%%%%%%%%%%%%%%%%%%%%%%%%%%%%%%%%%%%%%%%%%%%%%%%%
\section*{Acknowledgments}
We thank Matthew Szydagis and Xun Chen for useful discussions about the 
LUX and PandaX-II experiments.  CQG and CHL are partially
 supported by National Center for Theoretical Sciences and MoST (MOST 104-2112-M-009-020-MY3), while DH 
  by the National Science Centre (Poland) research project, decision  DEC-2014/15/B/ST2/00108.
%\section*{References}

\end{document}